# A Tamm Plasmon-Porous GaN Distributed Bragg Reflector Cavity

Jon R. Pugh, Edmund Harbord, Andrei Sarua, Peter Fletcher, Ye Tian, Tao Wang and Martin J. Cryan

*Abstract*— This paper reports on design, measurement and optimisation of a Tamm plasmon metal-DBR cavity for use in the green part of the visible spectrum. It uses an optimised silver layer thickness and a porous DBR created using a novel electro-chemical etching technique. This device has applications in low cost lasers, photodetectors and photoconductive switches for the visible wavelength range.

## I. Introduction

Distributed Bragg reflectors (DBRs) are essential components for the development of optoelectronic devices that are formed from multiple layers of alternating materials with varying refractive index. Vertical-Cavity Surface-Emitting Lasers (VCSELs) traditionally consist of a lower and an upper DBR mirror, sandwiching an active/confinement layer containing multiple Quantum Well's (QWs). In this paper, we focus on the possibility of substituting the upper DBR mirror of a traditional VCSEL by using Tamm Plasmon confinement. Tamm plasmons are a type of optical surface mode, analogous to Tamm [1] electronic surface states that occur at the interface of a crystal. Tamm modes occur at the termination of a photonic crystal lattice when the termination is by a metallic layer [1, 2]. These can be tuned [2, 3, 5, 6] by controlling the thickness of the metal layer and/or the top stack of the spacer layer. In doing this, we have built upon existing research [7] to realise such a cavity in the technologically important GaN system. This has been challenging up to now due to the low refractive-index contrast between AlN and GaN and also the large lattice-mismatch between AlN and GaN, both making it difficult to achieve DBRs with a high reflectance and also to maintain a good crystal quality which is important for the subsequent growth of a micro-cavity.

Recent advances in the fabrication of lattice-mismatched DBRs based on GaN/nano-porous GaN DBRs have shown high reflectance, R>99.5% [8 ,9] due to much higher refractive-index contrast between GaN and nano-porous GaN. Figure 1 shows a schematic of the structure we are considering here, the higher refractive contrast makes the Tamm plasmon achievable in only 11-periods, many fewer than used in similar devices [10].

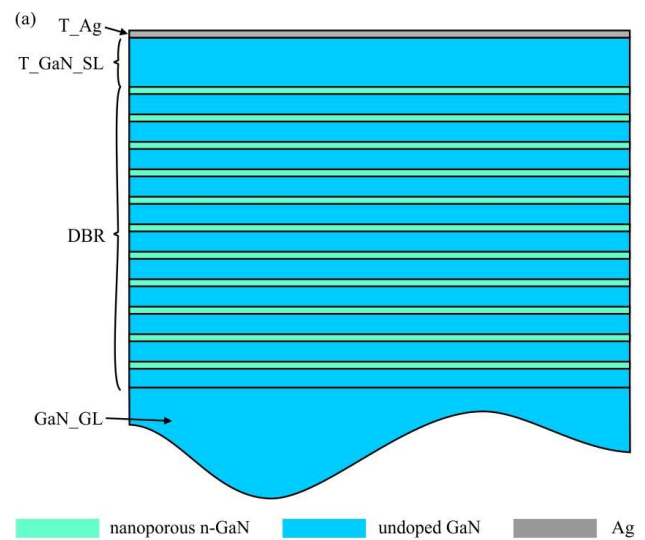

Fig. 1. (a) A schematic illustration of the epitaxially grown base structure on a sapphire substrate – this consists of a GaN growth layer (GaN_GL), 11-period DBR, GaN spacer layer (thickness=T_GaN_SL), and silver layer (thickness=T_Ag).

The simplicity of the Tamm structure in terms of growth time not only reduces the fabrication cost since only one DBR is required, it also allows for post growth tuning with the metal layer thickness. We hope this realization of a Tamm GaN cavity will spur the development of schemes for making short wavelength devices not only for VCSELs but also for photodetectors and photoconductive switches [11]. This technology could also be combined with emerging GaN integrated photonic devices [12, 13].

Manuscript received November 29, 2019; revised November ##, 2019; accepted November ##, 2019. Date of publication January ##, 2020. This work was supported by the Engineering and Physical Sciences Research Council through the Programmes Manufacturing of nano-engineered III-nitride semiconductors under grant EP/M015181/1 and Future Compound Semiconductor Manufacturing Hub under grant EP/P006973/1.

J. R. Pugh and M. J. Cryan are in the Photonics and Quantum Group within the Department of Electrical and Electronic Engineering, Merchant Venturers Building, Woodland Road, University of Bristol Bristol, BS8 1UB, U.K. (e-mail: jon.pugh@bristol.ac.uk; m.cryan@bristol.ac.uk).

A. Sarua is in the School of Physics, University of Bristol, HH Wills Physics Laboratory, Tyndall Avenue, Bristol, BS8 1TL, U.K. (email: a.sarua@bristol.ac.uk).

P. Fletcher, Y. Tian and T. Wang are in the Centre for GaN Materials and Devices within the Department of Electrical Engineering, University of Sheffield, 3 Solly Street, Sheffield, S1 4DE, U.K. (email: psfletcher1@sheffield.ac.uk; ytian18@sheffield.ac.uk; t.wang@sheffield.ac.uk).

## II. FABRICATION OF POROUS GaN DBR

The sample is grown on (0001) sapphire substrates using a low-pressure metal-organic-vapour phase epitaxy (MOVPE) system by means of a standard two-step growth method. After the substrate is initially subjected to thermal cleaning in flowing $H_2$, a low-temperature GaN nucleation layer with a thickness of 25 nm is prepared, followed by a 1.5 μm GaN layer grown at $1100^0C$. Subsequently, 11 pairs of undoped GaN and heavily silicon doped n-GaN are then grown, where the thicknesses of the undoped and the n-GaN in each pair are 50 nm and 65 nm, respectively, and the doping level is on a scale of $10^{19}$ cm$^{-3}$. A microcavity structure is then grown, and it consists of a 54 nm undoped GaN layer, 3 pairs of InGaN/GaN multiple quantum wells (MQW) and a further 54 nm undoped GaN layer. For the InGaN MQWs, the thicknesses of the InGaN quantum well and GaN barrier are 2.5 nm and 10 nm, respectively, and the indium content is ~20%. Finally, the 11 pairs of undoped and n-GaN layers are formed into nanoporous Distributed Bragg reflectors (DBRs) by an electro-chemical (EC) etching technique – similar procedures have been proposed by Zhang *et al* [8] and Zhu *et al* [9].

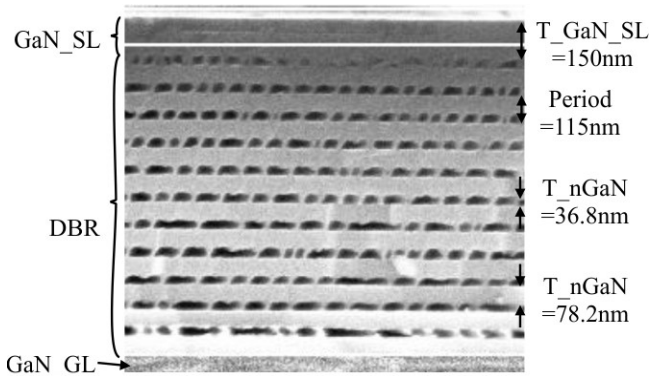

Fig. 2. An SEM image of the etched porous layers.

The mechanism of EC etching is based on a combination of an oxidation process and then a dissolution process in acidic solution under an anodic bias. Under a positive anodic bias, the injection of holes leads to the oxidation of GaN, and the oxidized layer is then chemically dissolved in an acidic electrolyte. Therefore, EC etching can only be performed on n-type GaN with high conductivity. In this case, n-GaN can be etching into a nanoporous structure whose refractive index depends on porosity, while undoped GaN remains unetched. Our EC processes are carried out in nitric acid with a mole concentration of 0.5 M as an electrolyte at room temperature, where n-doped GaN is used as an anode and a platinum foil as a cathode. The EC porosification process is carried out with a DC bias of 9V bias until the etching current drops to zero, which means that all the n-GaN layers have been etched into nanoporous GaN layers. Not all of the n-GaN was converted porous during the EC etching process (further investigation is on-going), so in the simulations that follow a ratio ($r$) of 0.32 was used for porous (36.8nm)/non-porous (78.2nm) layers, which was measured from the SEM image shown in fig. 2. The etched layer morphology is on the boundary of mesoporous/macroporous given the large width of the pores. Following the EC etching, firstly a 70nm thick layer of Ag has been evaporated onto the surface of the GaN spacer layer.

## III. SIMULATION AND MEASUREMENT OF OPTICAL BEHAVIOUR

A two-dimensional finite difference time domain (FDTD) approach has been used to simulate the multilayer structure using commercial-grade software [Lumerical Solutions, Inc. http://www.lumerical.com/tcad-products/fdtd/]. A broadband plane-wave source is incident onto the top of the structure, first passing through the 70nm thick Ag layer before impacting the 135nm thick GaN spacer layer and DBR structure – period ($a$) 115nm (11-periods). Beneath the DBR, there is a 1.6μm thick GaN growth layer on a Sapphire substrate. The pores are modelled by adding blocks of air of random width to approximate the dimensions shown in figure 2 – the thicknesses of nanoporous n-GaN and undoped GaN are 36.8nm and 78.2nm respectively. Given their thin vertical dimension, the QW layers are deemed to have negligible effect on the reflectance and are not added to the simulation structure. The refractive index for the all GaN material was obtained from Filmetrics [https://www.filmetrics.com/refractive-index-database/GaN/Gallium-Nitride], who cite Adachi [14] as the source and the "Ag (Silver) - CRC" material data within Lumerical is fitted to our spectral region of interest. The grown sample has four regions of interest mapped out in figure 3(d). There are two distinct areas, one with the EC etched DBR and one unetched. Both areas are partially masked and 70nm of Ag being evaporated onto the surface, gives four regions on the sample shown in the inset. A direct wide-angle measurement of the four regions is carried out using Fourier Image Spectroscopy (FIS) [15]. A white-light source is incident on the sample surface through, and collected by, the same high numerical-aperture (NA=0.8) 100x magnification microscope objective. The back focal-plane of this objective is scanned, which allows full angular reflection data from the sample, normalised to that of a silver mirror, to be plotted against wavelength.

The normal incidence reflectance from these four regions is captured between λ=400-620nm and are plotted using dashed lines for FDTD simulations and solid lines for FIS measurements in figure 3. The unetched/uncoated region (No DBR/No Ag – plot (d)) shows a rippled reflectance centred at ~0.2 with a ripple period between 10-20nm depending on spectral position. This is consistent with what one would expect for constructive/destructive interference between the GaN/air and GaN/Sapphire boundaries. When coated with Ag (No DBR+Ag – plot (b)), the reflectance increases due to the high reflectance of Ag, but due to its thin-film nature it is still relatively transparent at these wavelengths – the ripple period however does not change. Large reflectance above 0.99 is predicted in simulations from the DBR (DBR/No Ag – plot (c)) with a reflectance bandgap ranging from λ=432-553nm (FWHM). The porous regions are approximated to be perfect polygons in the FDTD simulations, therefore there is near-ideal Bragg reflection from the multiple layers. Irregular porous shapes introduce scattering, so the etched structure is no longer a perfect Bragg reflector and the overall peak measured reflection drops to ~0.88 in the measurements. A further important factor from the measurement is that the illumination covers a large angle even though we are only considering the reflected light normal to the surface. Given the DBR only reflects normally incident light near-perfectly, the fact that a large proportion of the incident light is off normal axis will cause a broadband drop in reflection. When coated with 70nm of Ag (DBR+Ag – plot Fig.3 (a)) a small dip is seen in the reflectance at around 505 nm caused by Tamm mode confinement at this wavelength. The strong increase in reflectance caused by the TAMM mode is present at all angles collected by the high-NA objective (not shown), with only very small variations in bandwidth and intensity. There is a blue-shift of approximately 30nm in the TAMM mode wavelength at the highest collection angles ($\theta \approx 53°$, where normal reflectance is $\theta = 0°$) caused by a change in the layer thicknesses seen by the light at these higher angles. We believe that the results would be the same for a smaller-magnification/lower-NA objective that would illuminate a significantly larger area. Our FDTD modelling predicted that by moving to T_Ag=24nm, a much larger dip would be obtained. The 70nm Ag was removed and replaced by a thinner layer which we estimated to be in the region of 24nm. Fig 3(a) shows that a good level of agreement is obtained between the measured and modelled cases. We believe the differences are primarily caused by the scattering discussed above.

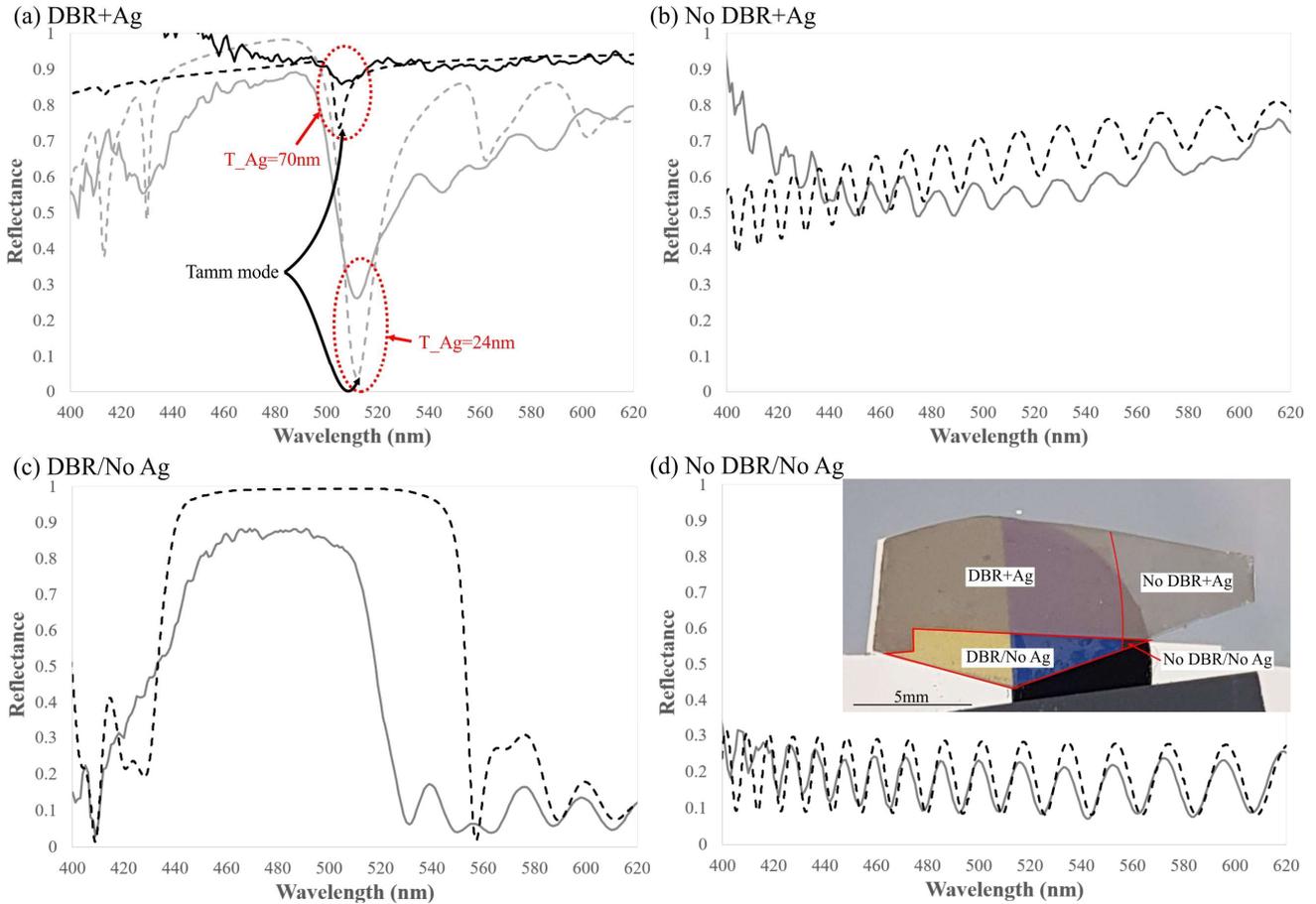

Fig. 3. FDTD simulated (dashed lines) and FIS measured normal to the interface (solid lines) reflectance from the four regions of the sample – a photo of which is shown in the inset – (a) DBR+Ag (24nm and 70nm thicknesses), (b) No DBR+Ag (24nm thickness), (c) DBR/No Ag, (d) No DBR/No Ag.

IV. CONCLUSION

We have demonstrated through numerical modelling and fabrication a Tamm mode resonance in a silver coated GaN/Porous GaN Bragg reflector structure. In future, careful design, that overlaps quantum wells (QWs) spatially with the regions of highest electric field intensity and spectrally with the resonance will enable novel LED and laser devices to be fabricated.